\documentclass[twocolumn,showpacs,preprintnumbers,amsmath,amssymb,prX]{revtex4}

\usepackage{graphicx}
\usepackage{dcolumn}
\usepackage{bm}

\begin{document}

\preprint{APS/123-QED}

\title{Community Detection via Facility Location}

\author{Jonathan W. Berry}
\author{Bruce Hendrickson}%
\author{Randall A. LaViolette}%
\author{Vitus J. Leung}%
\author{Cynthia A. Phillips}%
\affiliation{ 
Sandia National Laboratories, P.O. Box 5800, Albuquerque, NM, 87185.
}

\date{\today}

\begin{abstract}
In this paper we apply theoretical and practical results from facility location theory
to the problem of community detection in networks.  The result is an
algorithm that computes bounds on a minimization variant of local
modularity.  We also define the concept of an edge support and a
new measure of the goodness of community structures with respect to
this concept.  We present preliminary results and note that our
methods are massively parallelizable.
\end{abstract}

\pacs{02.10.Ox 02.60.Pn 89.75.Fb 89.75Hc}
\keywords{Community detection facility location modularity graph network}
\maketitle

\section{Introduction}
\label{sec:intro}
In this paper, we apply results from facility location theory to
community detection.  Leveraging recent developments in both fields,
we compute a weighting of the input graph that represents pertinent
information for community detection algorithms.  We show how to
compute this weighting efficiently using techniques from facility
location theory.  We can interpret the weights as probabilities and
randomly sample over a space of good community assignments.  Computing
the weights involves solving a linear program (LP)~\cite{HillierL95}
that has special structure.  Solvers for this special kind of LP
require only linear space and linear time per iteration.
Furthermore, this solution strategy is amenable to massive
parallelism.

We also give new measures for evaluating the quality of 
community assignments and show that our algorithms provide a provable
lower bound on solution quality with respect to one of these.  We
demonstrate empirically that another of our measures is complementary to 
modularity, and that optimizing based on this new measure better resolves
small communities in large graphs and better matches common sense 
community structures in familiar datasets.
Thus, we make four contributions in this work: 
we demonstrate a connection between community detection and facility location;
we use that connection to compute lower bounds on solution quality;
we show how to compute new measures for the goodness of	community structure that contrast with modularity; and
we apply massively parallelizable methods to compute these bounds and measures.

\section{Background}

Newman and Girvan's concept of {\em modularity}~\cite{ng2004} is now
ubiquitous in the community detection literature.  There are several
variations on this concept, such as~\cite{CapocciSCC05, FanLZWD06,
GfellerCD05, mrc2005, Zakharov07}, and many heuristics to optimize the
original concept and these variations,
e.g.~\cite{rb2004}\cite{cnm2004}.  In order to compute community
structures with good modularity in large network instances,
researchers commonly use one of two approaches: greedy heuristics,
such as~\cite{cnm2004} and~\cite{wt2007}, and metaheuristic
approaches, such as simulated annealing~\cite{rb2004}. Agarwal and
Kempe~\cite{ak2007} applied mathematical programming to the problem of
maximizing modularity, resulting in an algorithm to compute upper
bounds for that measure.

We present an alternative that employs results from the vast facility
location literature to community detection. We model a variation of modularity as an
{\em uncapacitated facility location problem} (to be defined below),
and employ the simple and powerful {\em Volume
algorithm}~\cite{ba2000} to solve the problem.  Mulvey and
Crowder~\cite{MulveyC79} used similar techniques, applying older
subgradient methods, to solve p-median problems that approximately
cluster points in $n$-dimensional space.

We first observe that specializing a minimization version of
the modularity problem produces an uncapacitated facility location problem.
We then discuss its solutions and the interpretation and use of its results. 

\section{Strongly-Local Modularity (SLM)}

Girvan and Newman define the {\em modularity} ($Q$) for a graph $G$ 
as follows: $Q = \sum_s (e_{ss} - a_s^2)$,
where $s$ is a community in the domain $\{1 \ldots q\}$, 
$e_{rs}$ is the fraction of $E(G)$ (the edge set of the graph) 
that connects a node in community $r$ to one in community $s$,
and $a_s$ is the fraction of edges that have at least one endpoint in
$s$ ($a_s = \sum_r e_{rs}$).  Squaring $a_s$ gives the probability that
an edge would have both endpoints in community $s$ in a random
graph with the same endpoint degree distribution. 
Modularity is a way to measure the quality
of community assignment: it rewards communities that are better connected 
than would be expected in a random graph reflecting the endpoint degree 
distribution.

Now consider a simple variation of the modularity concept:
$Q^- = \sum_s (1 - (e_{ss} - a_s^2)).$
Minimizing $Q^-$ is similar to, though not identical to, maximizing $Q$.
Basic algebra shows that a community assignment minimizing 
$Q^-$ has at most as many communities as one that maximizes $Q$, and
this is typically a strict inequality.
%
%

It is well-known that community assigments of maximum modularity
fail to resolve small communities in large graphs~\cite{fb2007}.  Reflecting
on this work, it would seem that the $Q^-$ measure will compound this problem
by resolving even fewer communities.  However, we provide a remedy via
a further modification described below, and our switching of optimization sense 
will prove useful.

Muff, Rao, and Caflisch~\cite{mrc2005} define
the {\em local modularity} to be the same as modularity, except that
the denominators in the fractions $e_{rs}$ are the numbers of edges
in a cluster's ``neighborhood,'' defined to be itself and all neighboring
clusters.  We use a metric that also focuses on local structure, but is even
more restrictive, requiring no information about the structure of neighboring
communities.
We define a {\em strongly local community}
to consist of a single representative node and all of its
immediate neighbors, i.e., a full community of radius one.  
Let $Q_s = e_{ss} - a_s^2$.  We can compute this measure
for any strongly local community without
knowing any community assignments other than the vertices in $s$.  
Ignoring algorithmic details, we need only know the
number of triangles in the strongly local community and the degree of 
each node.

Now we give the key definition that allows us to model the problem using
facility location theory.  Let
\[ \tilde{Q}_s = \left\{ \begin{array}{ll}
			Q_s  & \mbox{if $s$ is a strongly local community} \\
			0      & \mbox{otherwise} \\
		     \end{array}
	     \right.\]

We define the {\em Strongly-Local Modularity} (SLM) as follows:
\[\tilde{Q}^- = \sum_s (1 - \tilde{Q}_s).\]
We use SLM in combination with a relaxed notion of community assignment 
in which community representatives can share common neighbors within
their respective communities.

\section{Modeling SLM as a facility location problem}
\label{sec:fac}

We transform instances of the community detection problem into instances of 
the {\em Uncapacitated Facility Location Problem} (UFLP)\cite{HillierL95}.  Given a set
of potential facility locations $L$, a set of customers $C$, a set of
facility opening costs $f_i$, and a set of service costs $c_{ij}$ (the
cost to serve customer $j$ using facility $i$), the objective function of UFLP
is
\[F(x) = \sum_{i \in L} f_i x_i + \sum_{i \in L, j \in C} c_{ij} y_{ij},\]
where the variables $x_i$ indicate whether or not location $i$ hosts a facility,
and the variables $y_{ij}$ indicate whether or not location $i$ serves 
customer $j$.  Solutions to UFLP minimize $F(x)$ subject to the constraints
that every customer must be served, and that no customer can be served by a 
facility that does not exist.  UFLP is a
well known NP-hard problem~\cite{CornuejolsNW90,GareyJ79, GuhaK99}, yet it has special structure that enables
efficient computations of fractional solutions.

We consider all vertices to be potential facility locations, with
facility opening costs $f_s = (1 - \tilde{Q}_s)$.  Each vertex is a
customer that must be served by a facility (and may serve itself if it
hosts a facility).  The service cost is zero for a node to serve a
neighbor in the graph.  Nodes cannot serve non-neighbors (cost is
effectively infinite).  The solution to the UFLP is a minimum-cost
facility and service assignment in which every vertex is served.

UFLP is an integer program (IP), but we need only solve the linear
programming relaxation of the IP\cite{HillierL95}.  This relaxation
has special structure that obviates the need for a general linear
program solver. We apply Lagrangian relaxation in conjunction with an
elegant subgradient method known as the Volume algorithm
(VA)~\cite{ba2000} in the Lagrangian relaxation framework
of~\cite{bc2005}.  The memory usage of this combined procedure is on
the order of the problem input size.  VA makes a series of
linear-time passes over the data.  There are no known asymptotic bounds
on the number of iterations. However, in practice, the total runtime
is comparable to the $O(n \log^2 n)$ runtime of the most familiar fast
modularity heuristic, the {\em CNM} greedy
algorithm~\cite{cnm2004}. We have observed this experimentally on
graphs with up to 100 million edges.

The volume algorithm provides a fractional solution to the UFLP that in turn
provides a provable lower bound on $\tilde{Q}^-$ where all communities are
strongly local. 

\begin{figure}[tbh]
\centerline{\includegraphics*[height=2in]{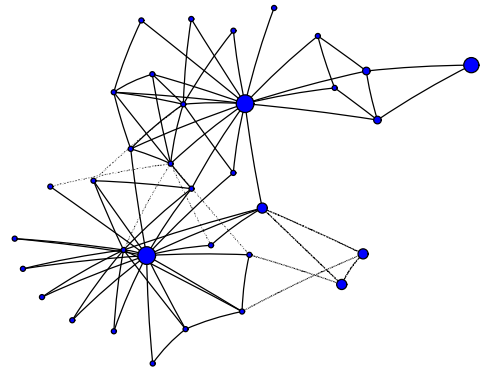}}
\caption{The support of Zachary's karate club.  Solid edges have stronger
	support than speckled edges and larger vertices are more likely to
	be leaders.  Note the nearly-invisible edges linking 
	portions of the club destined to split. 
		\label{fig:support}}
\end{figure}

Our community-assignment procedure selects a set of facilities to
``open.''  Each open facility represents a leader of a subset of a
strongly-local community.  That is, every community has at least one
node that is adjacent to all other nodes in the community.  The set
of leaders, therefore, forms a {\em dominating set}, that is, a set of
vertices $D$ such that each vertex in the graph is either in $D$ or
adjacent to an element of $D$.
\begin{figure*}[tbh]
\begin{tabular}{ccc} 
\includegraphics*[height=1.00in]{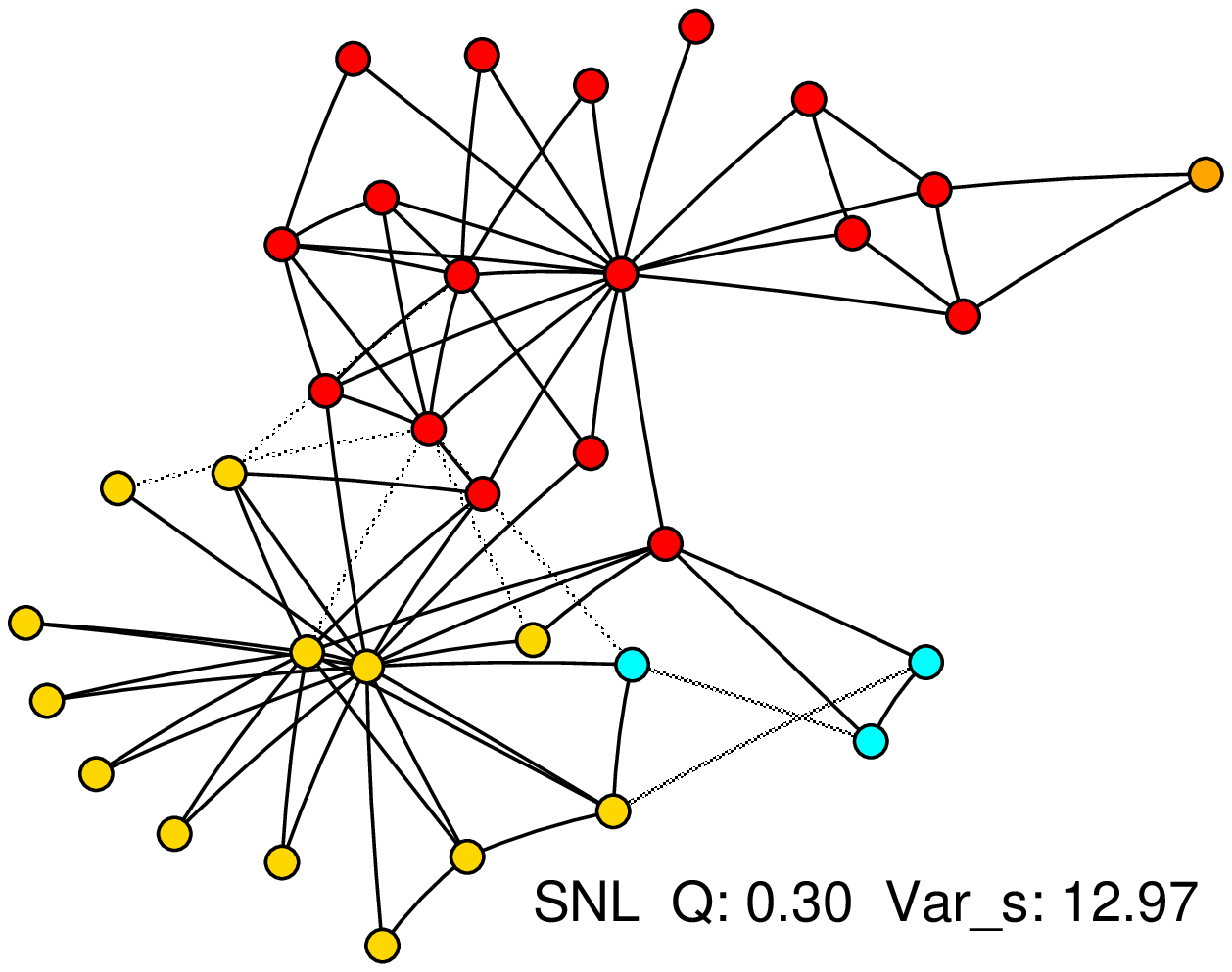} & &
\includegraphics*[height=1.00in]{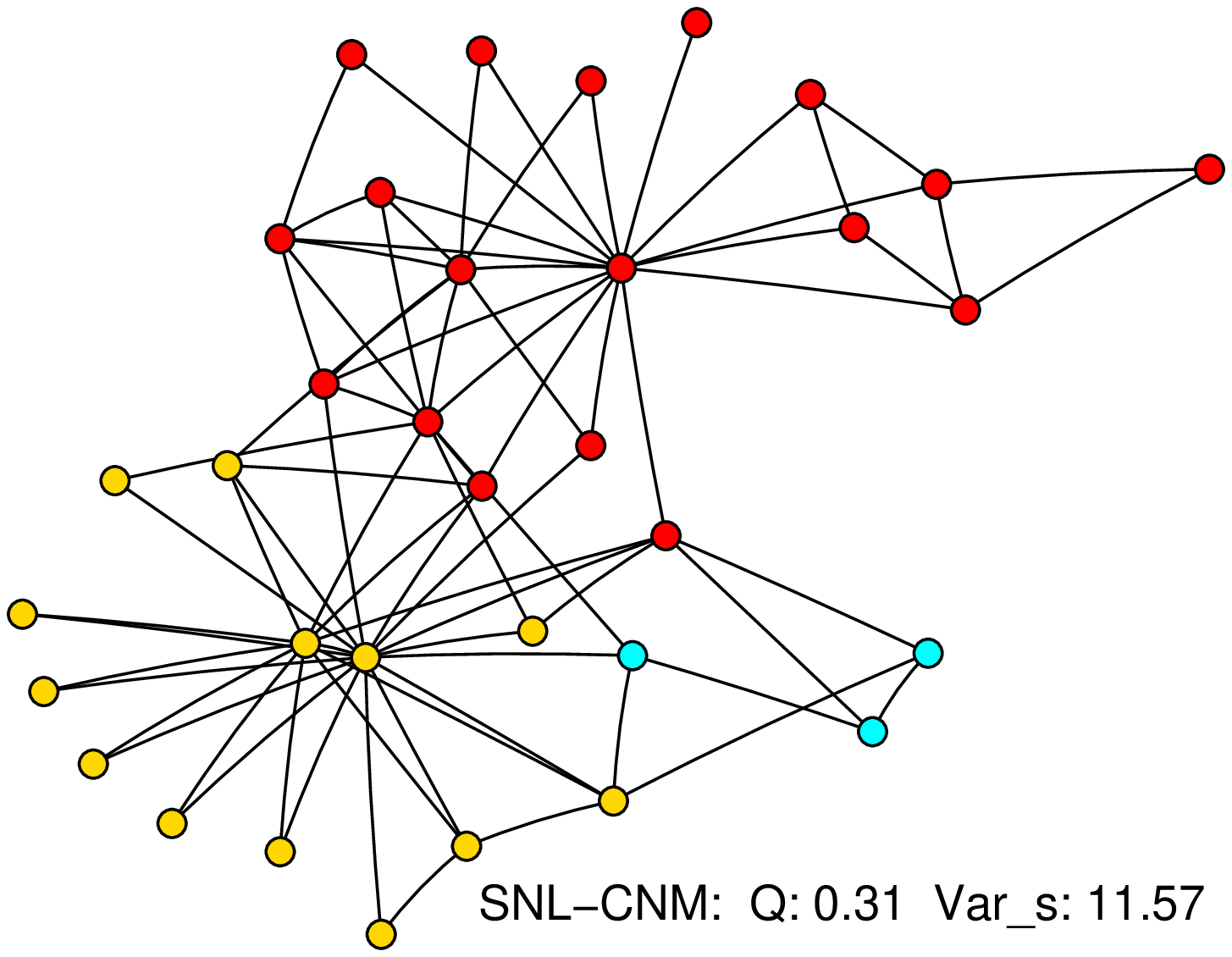} \\
 & \includegraphics*[height=1.00in]{figures/support.ps} & \\
\includegraphics*[height=1.00in]{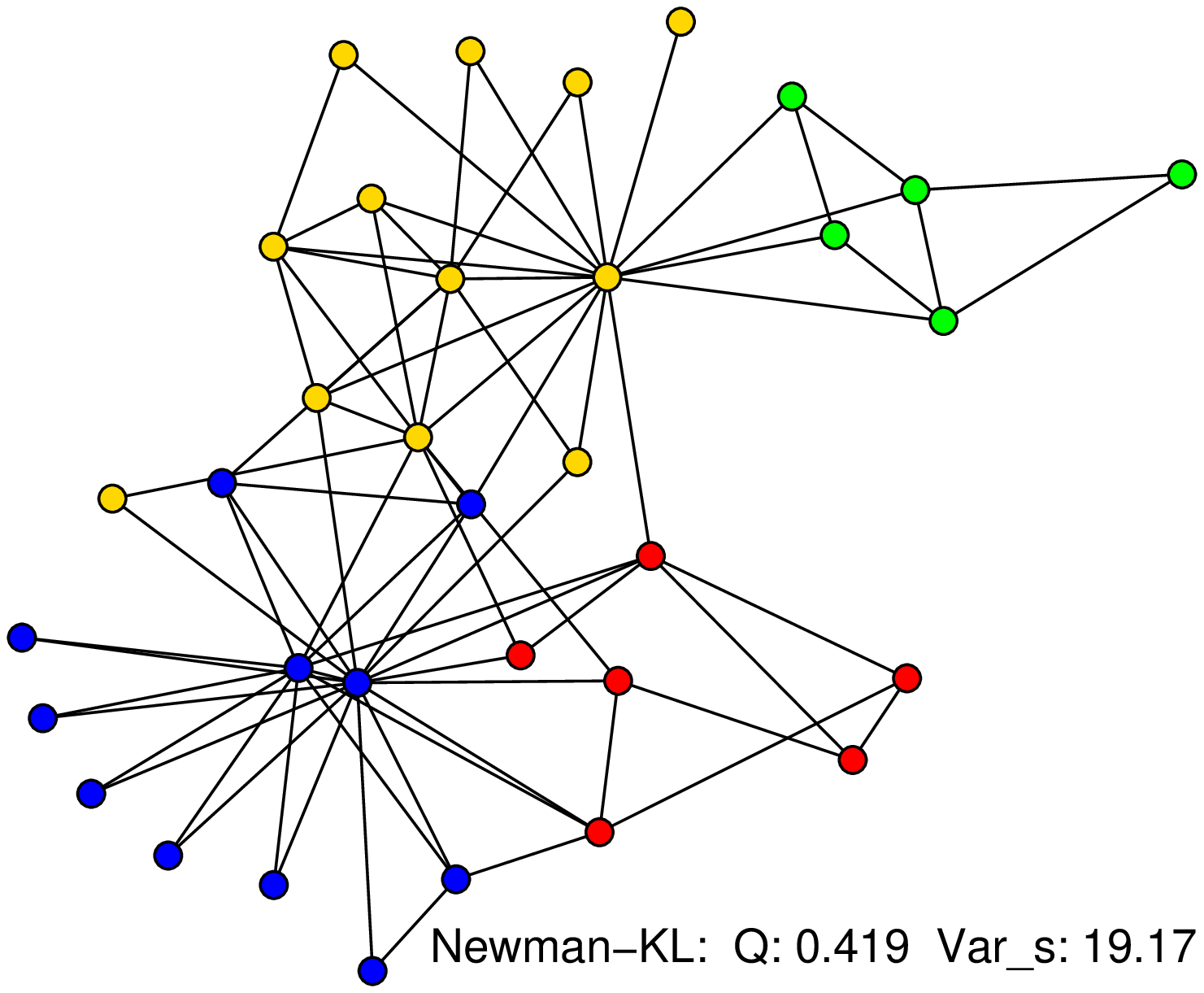} & &
\includegraphics*[height=1.00in]{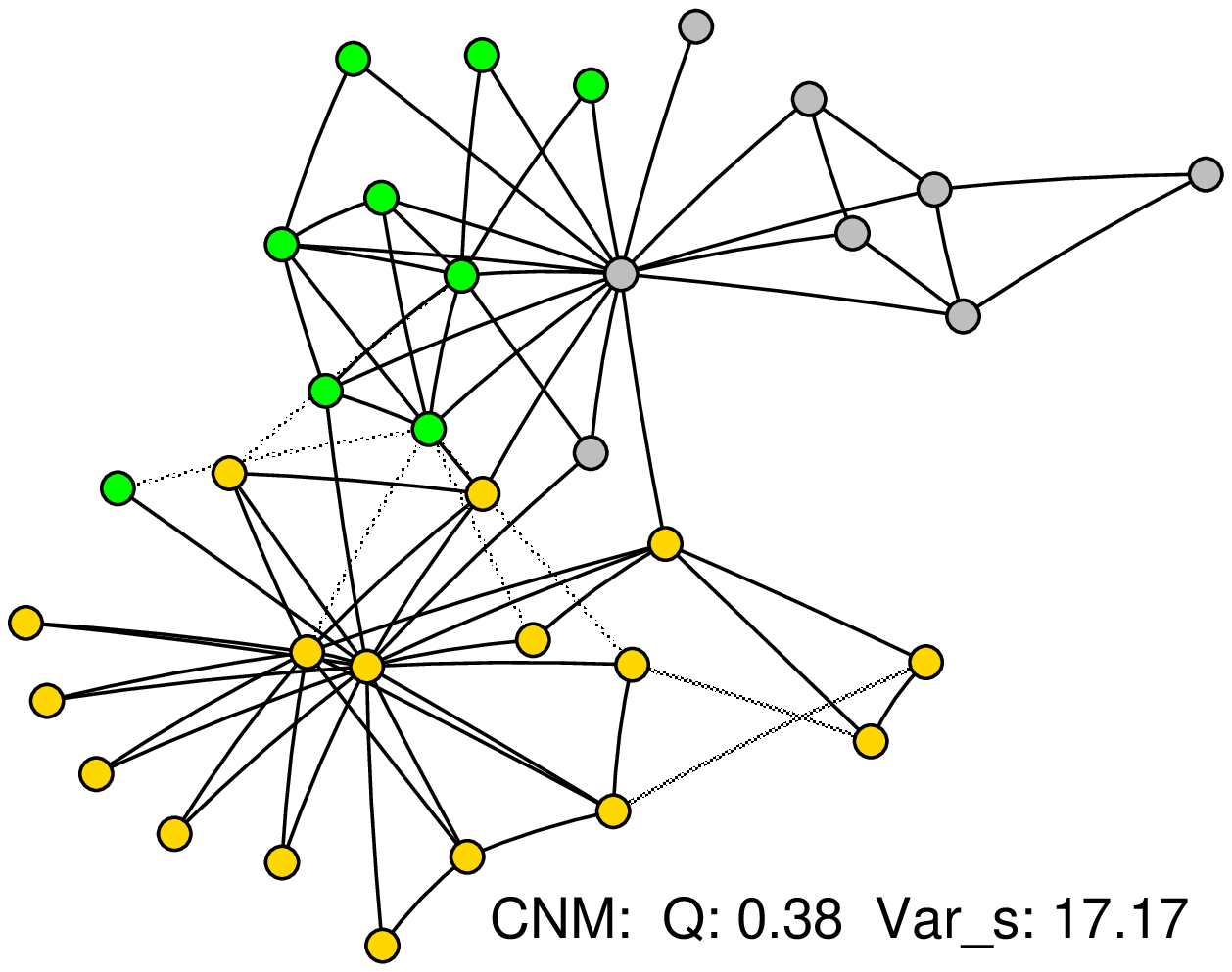}
\end{tabular}
\caption{The support of Zachary's karate club and its relationship to
	 actual solutions of various algorithms.  The 
	support variance $\mbox{Var}_s$ decreases as solutions agree
	more closely with the support.  Note that 
	the community assignments with maximum modularity split edges
	with strong support within both of the true communities.
	\label{fig:support-algs}}
\end{figure*}

In our community-finding procedure, called {\em SNL}, we set the
facility-opening costs as described above and use VA to compute an
optimal fractional placement of facilities.  We then open each
facility with probability equal to its fractional assignment value.
If this does not produce a dominating set, then we repair it to make a
dominating set.  We then assign all the other vertices to a community.
There are a number of ways one can do this.  In this paper, we assign
each non-selected node to the selected neighbor with highest fractional
facility placement.

\section{The Support}

We define the {\em support} of an edge $(u,v)$ to be a real number between
$0$ and $1$ that indicates the level of support/evidence for nodes $u$
and $v$ being in the same community.  Given any randomized algorithm
$A$ for community detection, such as the metaheuristic approach
of~\cite{rb2004}, we can compute a support with respect to $A$ by
sampling: generate many community assignments using $A$, then compute
the fraction of times each edge is intra-community. We now show how to
compute a support with respect to {\em SNL} without sampling.

Given a fractional solution $x$ to an instance of UFLP, we define the 
support with respect to {\em SNL} 
to be a set of values $z$, where $z_j$ 
is a probability that in a set of community leaders sampled from $x$, 
edge $j$ could link two 
vertices in the same community.  Formally, 
\[z_{e=(v,w)} = 1 - [(1-x_v) * (1-x_w) * \Pi_{u \in N(v) \cap N(w)} (1-x_u)].\]

An edge $e = (v,w)$ has strong support if it is unlikely that {\em none} of 
the vertices capable of serving both $v$ and $w$ will become a server.
This includes $v$, $w$, and their mutual neighbors.
Figure~\ref{fig:support} depicts the support of Zachary's karate club
dataset~\cite{z1977}, an abstraction of a social network that famously split into two.  The larger vertices and darker edges 
have higher $x$ and $z$ values, respectively.  Even before community
assignments have been specified, the community structure begins to
emerge in fractional form.  Note that some edges that are destined to
become inter-community edges have very low support and are therefore
almost invisible.

Given the support of a graph, we define a new measure to evaluate the 
effectiveness of community assigments. We define the {\em support variance}
($\mbox{Var}_s$) as follows, assuming that $\delta(v,w)$ is
an indicator function with value 1 if $v$ and $w$ are in the same community
and 0 otherwise.
\[  \mbox{Var}_s = \sum_{(v,w) \in E(G)} (\delta(v,w) - y_{vw})^2. \]

\begin{figure*}[tbh]
\centerline{\includegraphics*[height=1.0in]{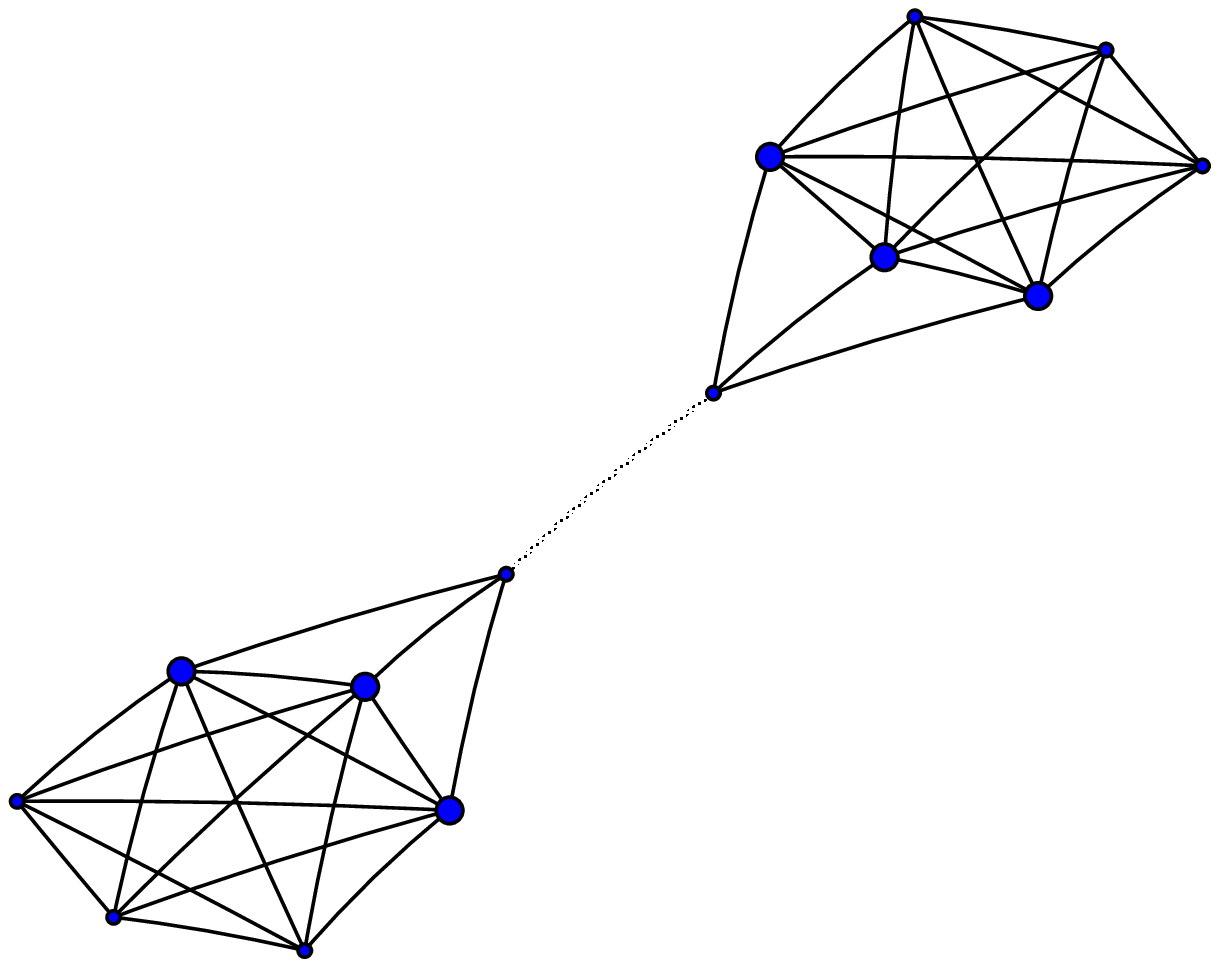}
	    \includegraphics*[height=1.0in]{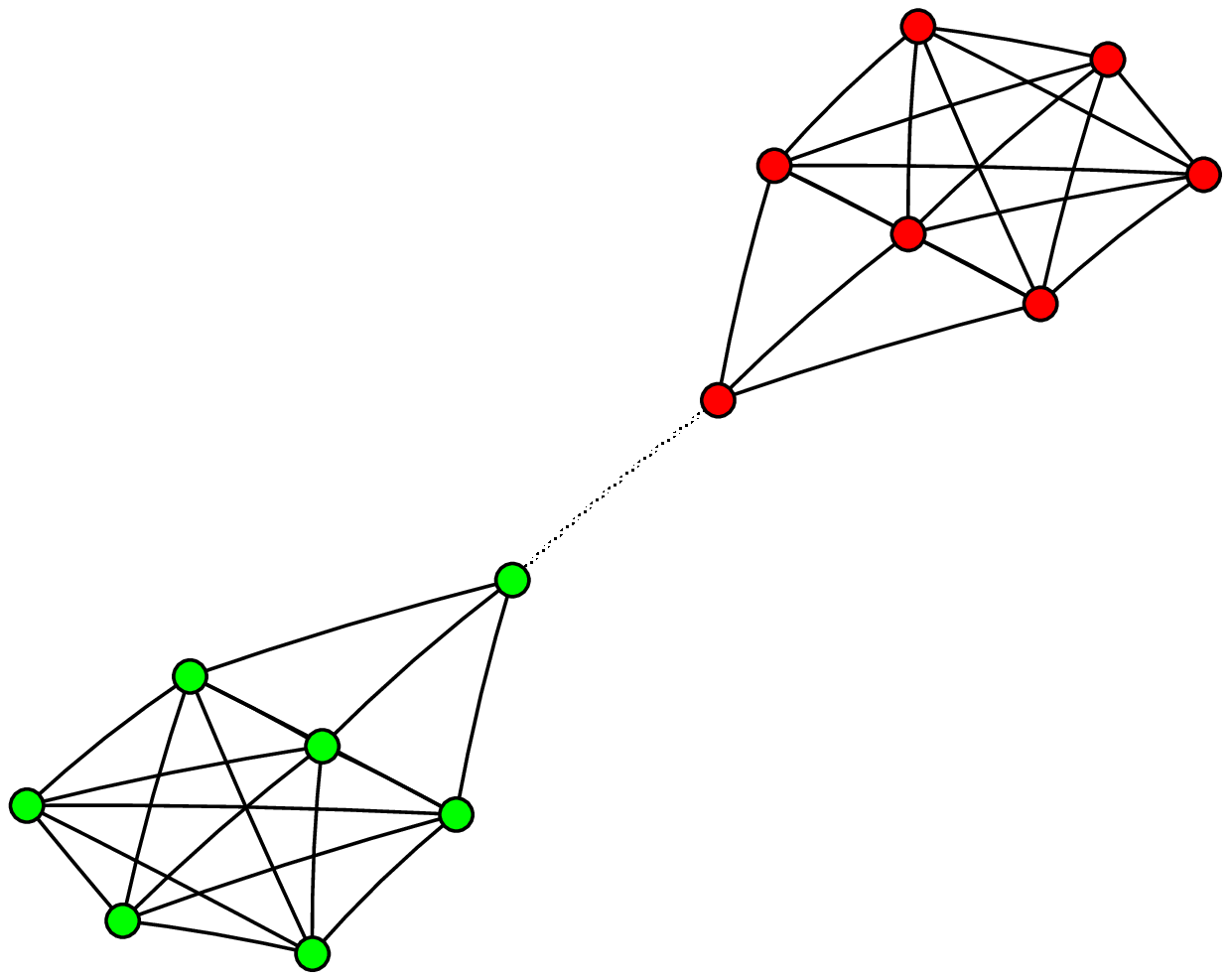}
	\hspace*{0.5in}
	    \includegraphics*[height=1.0in]{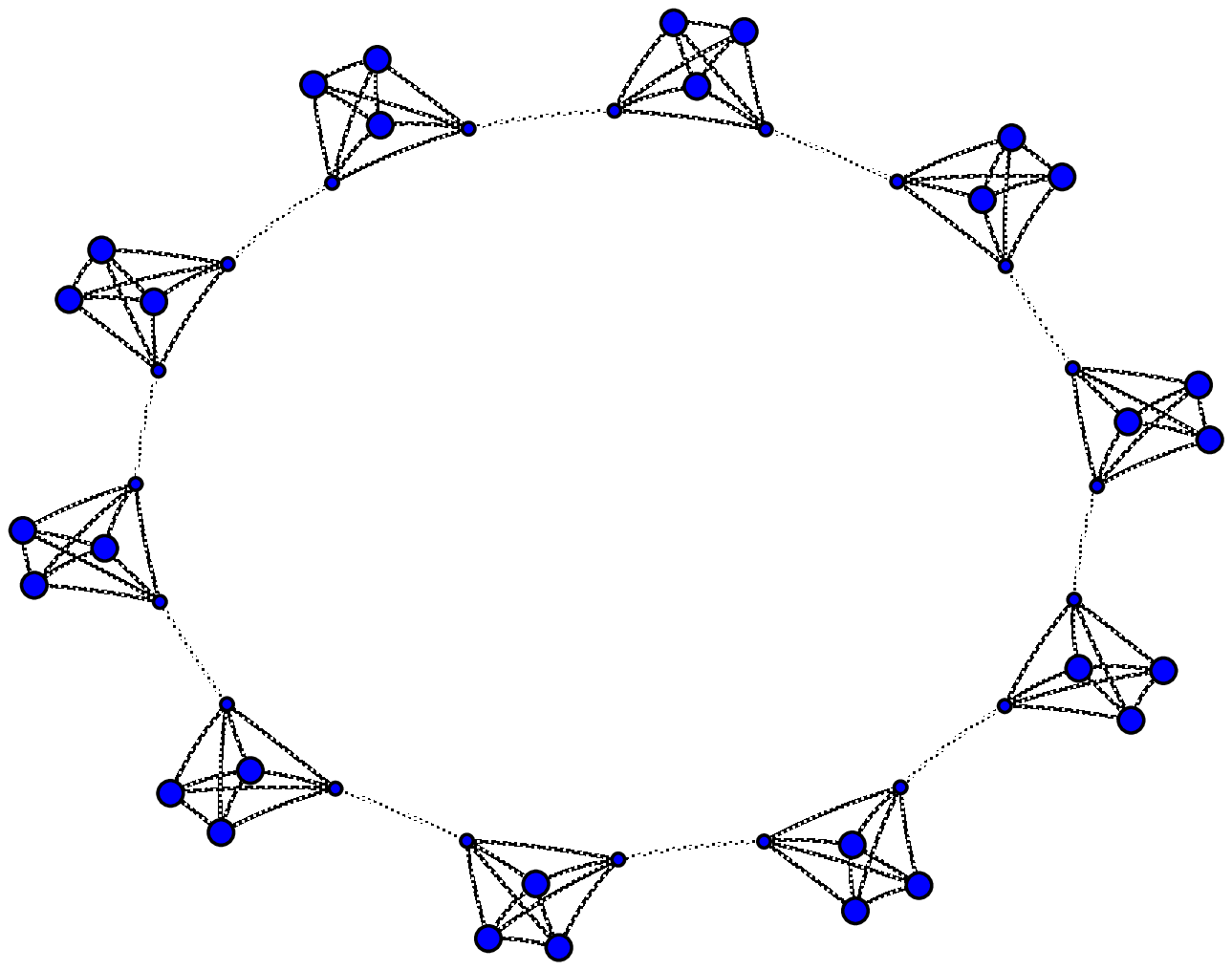}
	    \includegraphics*[height=1.0in]{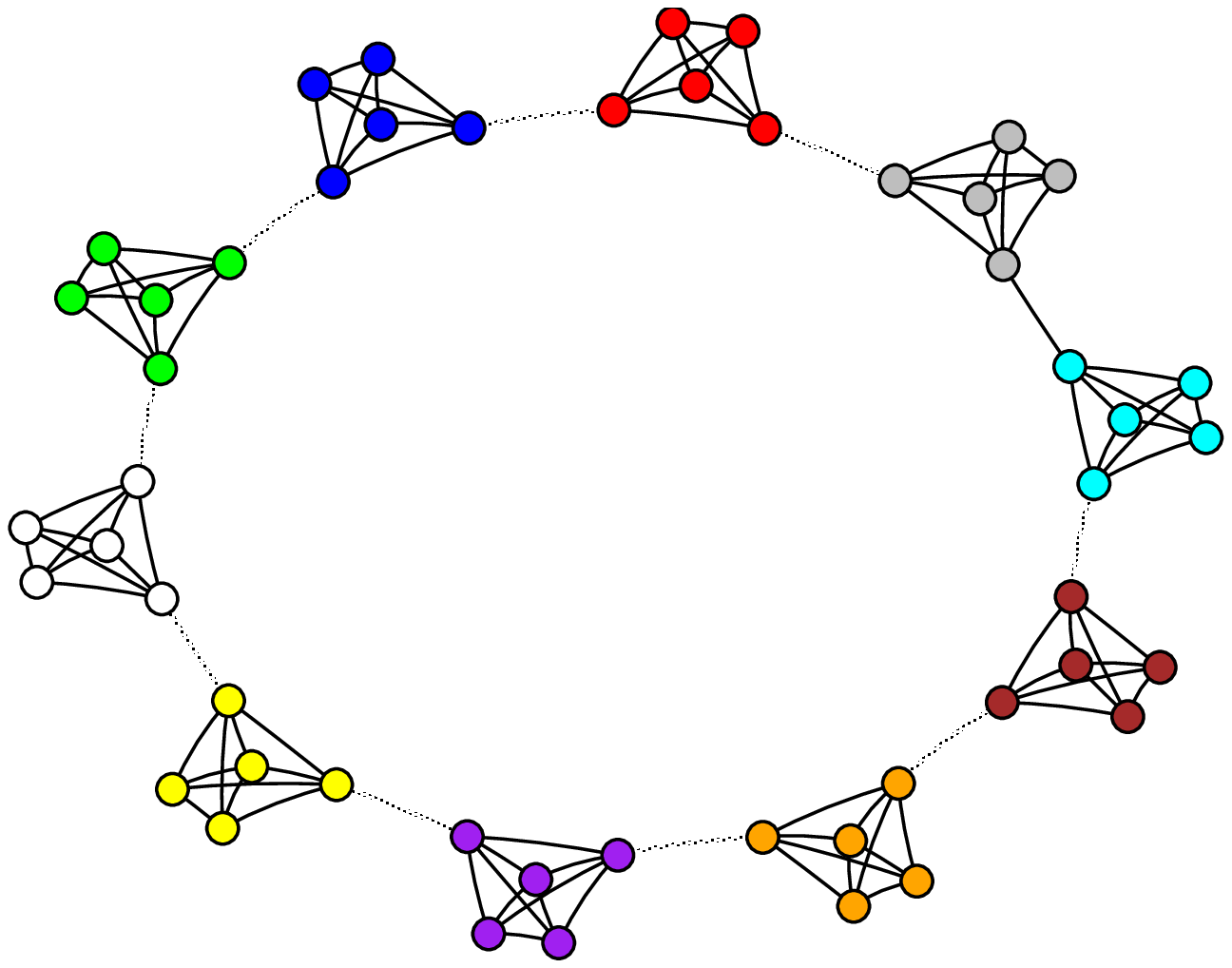}}
\caption{Maximizing modularity on these instances is known to produce
	non-intuitive answers.  However, each instance has a support
	that agrees with common sense and leads to intuitive rounded
	solutions.  The left hand
	instance is from\protect{~\cite{rb2004}}, and the right 
	instance is the ring-of-cliques example from\protect{\cite{fb2007}}.
	As more cliques are added to the ring, modularity optimization
	will merge cliques, increasing the support variance.  The facility
	location-based solution is not sensitive to the number of cliques.
	\label{fig:tough}}
\end{figure*}

\section{Preliminary Computational Results}

We applied our methods to several familiar
datasets.  Figure~\ref{fig:support-algs} shows the support of
Zachary's karate club The colored images in
Figure~\ref{fig:support-algs} depict the solutions of four algorithms:
our facility location-based rounding heuristic ({\em SNL}); the {\em
CNM} greedy algorithm; a combination of these two ({\em SNL-CNM}), in
which {\em SNL} is used to compute strongly-local communities, then
{\em CNM} is allowed to merge these; and the eigenvector-based
approach of Newman, augmented with a Kernighan-Lin-like postprocessing
step ({\em Newman-KL})~\cite{n2006}.  {\em Newman-KL} gives one of the
best known values for modularity.

In this case, intuition and history favor the 
facility location-based community assigments with low support variance over
those with high modularity. For example, the latter split the topmost 
community despite reasonably strong support for the edges holding it together.

Figure~\ref{fig:tough} shows two instances that have been demonstrated in
recent literature to present inherent problems for modularity algorithms.
The modularity of the left hand instance, from~\cite{rb2004}, tricks 
greedy algorithms into merging the endpoints of the edge that has the least
support in their first step. The right hand instance, from~\cite{fb2007}, has
been used to show that modularity optimization fails to resolve small 
communities in large graphs. The example shown is a ring of ten 5-cliques,
and grouping the 5-cliques individually both minimizes support variance and
maximizes modularity.  However, as the number of 5 cliques increases, the
common sense solution continues to minimize support variance, but is discarded
by modularity optimization methods in favor of larger communities.

\section{Conclusions}

We have applied models and algorithms from facility location theory to
the problem of community detection, yielding an algorithm to compute a
provable lower bound on a minimization variant of local modularity, a
support measure that can be computed without sampling, and a
randomized rounding heuristic that can be generalized into a class of
heuristics.  We have also introduced a new measure for evaluating the
quality of community structures.  The effectiveness of our heuristics
for large graphs remains open, but the solution techniques themselves
are scalable and based upon simple traversals of the network that are
massively parallelizable in a more natural way than the priority
queue-based methods previously published.  We will explore the
scalability of our methods on supercomputers in future work.

\section{Acknowledgements}

We thank Aaron Clauset and Mark Newman for providing code and data for
our preliminary study.  We generated the images using Sandia's
informatics framework. Sandia is a multiprogram laboratory operated by
Sandia Corporation, a Lockheed Martin Company, for the United States
Department of Energy's National Nuclear Security Administration under
contract DE-AC04-94Al85000.  This work was funded by the Laboratory
Directed Research and Development program.

\bibliography{apssamp2}

\begin{thebibliography}{20}
\expandafter\ifx\csname natexlab\endcsname\relax\def\natexlab#1{#1}\fi
\expandafter\ifx\csname bibnamefont\endcsname\relax
  \def\bibnamefont#1{#1}\fi
\expandafter\ifx\csname bibfnamefont\endcsname\relax
  \def\bibfnamefont#1{#1}\fi
\expandafter\ifx\csname citenamefont\endcsname\relax
  \def\citenamefont#1{#1}\fi
\expandafter\ifx\csname url\endcsname\relax
  \def\url#1{\texttt{#1}}\fi
\expandafter\ifx\csname urlprefix\endcsname\relax\def\urlprefix{URL }\fi
\providecommand{\bibinfo}[2]{#2}
\providecommand{\eprint}[2][]{\url{#2}}

\bibitem[{\citenamefont{Agarwal and Kempe}(2007)}]{ak2007}
\bibinfo{author}{\bibnamefont{Agarwal}, \bibfnamefont{G.}}, and
  \bibinfo{author}{\bibfnamefont{D.}~\bibnamefont{Kempe}},
  \bibinfo{year}{2007}, \bibinfo{note}{arXiv:0710.2533v1}.

\bibitem[{\citenamefont{Barahona and Anbil}(2000)}]{ba2000}
\bibinfo{author}{\bibnamefont{Barahona}, \bibfnamefont{F.}}, and
  \bibinfo{author}{\bibfnamefont{R.}~\bibnamefont{Anbil}},
  \bibinfo{year}{2000}, \bibinfo{journal}{Mathematical Programming}
  \textbf{\bibinfo{volume}{87}}(\bibinfo{number}{0025-5610}).

\bibitem[{\citenamefont{Barahona and Chudak}(2005)}]{bc2005}
\bibinfo{author}{\bibnamefont{Barahona}, \bibfnamefont{F.}}, and
  \bibinfo{author}{\bibfnamefont{F.}~\bibnamefont{Chudak}},
  \bibinfo{year}{2005}, \bibinfo{journal}{Discrete Optimization}
  \textbf{\bibinfo{volume}{2}}(\bibinfo{number}{1}).

\bibitem[{\citenamefont{Capocci} \emph{et~al.}(2005)\citenamefont{Capocci,
  Servedio, Caldarelli, and Colaiori}}]{CapocciSCC05}
\bibinfo{author}{\bibnamefont{Capocci}, \bibfnamefont{A.}},
  \bibinfo{author}{\bibfnamefont{V.}~\bibnamefont{Servedio}},
  \bibinfo{author}{\bibfnamefont{G.}~\bibnamefont{Caldarelli}}, and
  \bibinfo{author}{\bibfnamefont{F.}~\bibnamefont{Colaiori}},
  \bibinfo{year}{2005}, \bibinfo{journal}{Physica A}
  \textbf{\bibinfo{volume}{352}}(\bibinfo{number}{2-4}), \bibinfo{pages}{669}.

\bibitem[{\citenamefont{Clauset} \emph{et~al.}(2004)\citenamefont{Clauset,
  Newman, and Moore}}]{cnm2004}
\bibinfo{author}{\bibnamefont{Clauset}, \bibfnamefont{A.}},
  \bibinfo{author}{\bibfnamefont{M.}~\bibnamefont{Newman}}, and
  \bibinfo{author}{\bibfnamefont{C.}~\bibnamefont{Moore}},
  \bibinfo{year}{2004}, \bibinfo{journal}{Phys.\ Rev.\ E}
  \textbf{\bibinfo{volume}{70}}(\bibinfo{number}{066111}).

\bibitem[{\citenamefont{Cornuejols}
  \emph{et~al.}(1990)\citenamefont{Cornuejols, Nemhauser, and
  Wolsey}}]{CornuejolsNW90}
\bibinfo{author}{\bibnamefont{Cornuejols}, \bibfnamefont{G.}},
  \bibinfo{author}{\bibfnamefont{G.~L.} \bibnamefont{Nemhauser}}, and
  \bibinfo{author}{\bibfnamefont{L.~A.} \bibnamefont{Wolsey}},
  \bibinfo{year}{1990}, in \emph{\bibinfo{booktitle}{{Discrete Location
  Theory}}}, edited by
  \bibinfo{editor}{\bibfnamefont{P.}~\bibnamefont{Mirchandani}} and
  \bibinfo{editor}{\bibfnamefont{R.}~\bibnamefont{Francis}}
  (\bibinfo{publisher}{John Wiley and Sons}, \bibinfo{address}{New York}), pp.
  \bibinfo{pages}{119--171}.

\bibitem[{\citenamefont{Fan} \emph{et~al.}(2006)\citenamefont{Fan, Li, Zhang,
  Wu, and Di}}]{FanLZWD06}
\bibinfo{author}{\bibnamefont{Fan}, \bibfnamefont{Y.}},
  \bibinfo{author}{\bibfnamefont{M.}~\bibnamefont{Li}},
  \bibinfo{author}{\bibfnamefont{P.}~\bibnamefont{Zhang}},
  \bibinfo{author}{\bibfnamefont{J.}~\bibnamefont{Wu}}, and
  \bibinfo{author}{\bibfnamefont{Z.}~\bibnamefont{Di}}, \bibinfo{year}{2006},
  \bibinfo{title}{The role of weight on community structure of networks},
  \bibinfo{note}{arXiv:physics/0609218}.

\bibitem[{\citenamefont{Fortunato and Barth\'{e}lemy}(2007)}]{fb2007}
\bibinfo{author}{\bibnamefont{Fortunato}, \bibfnamefont{S.}}, and
  \bibinfo{author}{\bibfnamefont{M.}~\bibnamefont{Barth\'{e}lemy}},
  \bibinfo{year}{2007}, \bibinfo{journal}{PNAS}
  \textbf{\bibinfo{volume}{104}}(\bibinfo{number}{1}), \bibinfo{pages}{36}.

\bibitem[{\citenamefont{Garey and Johnson}(1979)}]{GareyJ79}
\bibinfo{author}{\bibnamefont{Garey}, \bibfnamefont{M.}}, and
  \bibinfo{author}{\bibfnamefont{D.}~\bibnamefont{Johnson}},
  \bibinfo{year}{1979}, \emph{\bibinfo{title}{Computers and Intractability: {A}
  Guide to the Theory of {NP}-Completeness}} (\bibinfo{publisher}{Freeman, New
  York}).

\bibitem[{\citenamefont{Gfeller} \emph{et~al.}(2005)\citenamefont{Gfeller,
  Chappelier, and Rios}}]{GfellerCD05}
\bibinfo{author}{\bibnamefont{Gfeller}, \bibfnamefont{D.}},
  \bibinfo{author}{\bibfnamefont{J.}~\bibnamefont{Chappelier}}, and
  \bibinfo{author}{\bibfnamefont{P.~D.~L.} \bibnamefont{Rios}},
  \bibinfo{year}{2005}, \bibinfo{journal}{Physical Review E}
  \textbf{\bibinfo{volume}{72}}(\bibinfo{number}{5}), \bibinfo{pages}{056135}.

\bibitem[{\citenamefont{Guha and Khuller}(1999)}]{GuhaK99}
\bibinfo{author}{\bibnamefont{Guha}, \bibfnamefont{S.}}, and
  \bibinfo{author}{\bibfnamefont{S.}~\bibnamefont{Khuller}},
  \bibinfo{year}{1999}, \bibinfo{journal}{Journal of Algorithms} ,
  \bibinfo{pages}{228}.

\bibitem[{\citenamefont{Hillier and Lieberman}(1995)}]{HillierL95}
\bibinfo{author}{\bibnamefont{Hillier}, \bibfnamefont{F.~S.}}, and
  \bibinfo{author}{\bibfnamefont{G.~J.} \bibnamefont{Lieberman}},
  \bibinfo{year}{1995}, \emph{\bibinfo{title}{Introduction to Operations
  Research}} (\bibinfo{publisher}{McGraw-Hill}, \bibinfo{address}{New York,
  NY}).

\bibitem[{\citenamefont{Muff} \emph{et~al.}(2005)\citenamefont{Muff, Rao, and
  Caflisch}}]{mrc2005}
\bibinfo{author}{\bibnamefont{Muff}, \bibfnamefont{S.}},
  \bibinfo{author}{\bibfnamefont{F.}~\bibnamefont{Rao}}, and
  \bibinfo{author}{\bibfnamefont{A.}~\bibnamefont{Caflisch}},
  \bibinfo{year}{2005}, \bibinfo{journal}{Phys.\ Rev.\ E}
  \textbf{\bibinfo{volume}{72}}(\bibinfo{number}{056107}).

\bibitem[{\citenamefont{Mulvey and Crowder}(1979)}]{MulveyC79}
\bibinfo{author}{\bibnamefont{Mulvey}, \bibfnamefont{J.~M.}}, and
  \bibinfo{author}{\bibfnamefont{H.~P.} \bibnamefont{Crowder}},
  \bibinfo{year}{1979}, \bibinfo{journal}{Management Science}
  \textbf{\bibinfo{volume}{25}}(\bibinfo{number}{4}), \bibinfo{pages}{329}.

\bibitem[{\citenamefont{Newman}(2006)}]{n2006}
\bibinfo{author}{\bibnamefont{Newman}, \bibfnamefont{M.}},
  \bibinfo{year}{2006}, \bibinfo{journal}{PNAS}
  \textbf{\bibinfo{volume}{103}}(\bibinfo{number}{23}), \bibinfo{pages}{8577}.

\bibitem[{\citenamefont{Newman and Girvan}(2004)}]{ng2004}
\bibinfo{author}{\bibnamefont{Newman}, \bibfnamefont{M.}}, and
  \bibinfo{author}{\bibfnamefont{M.}~\bibnamefont{Girvan}},
  \bibinfo{year}{2004}, \bibinfo{journal}{Phys.\ Rev.\ E}
  \textbf{\bibinfo{volume}{69}}(\bibinfo{number}{026113}).

\bibitem[{\citenamefont{Reinhardt and Bornholdt}(2004)}]{rb2004}
\bibinfo{author}{\bibnamefont{Reinhardt}, \bibfnamefont{J.}}, and
  \bibinfo{author}{\bibfnamefont{S.}~\bibnamefont{Bornholdt}},
  \bibinfo{year}{2004}, \bibinfo{journal}{Phys.\ Rev.\ Lett.}
  \textbf{\bibinfo{volume}{93}}(\bibinfo{number}{218701}).

\bibitem[{\citenamefont{Wakita and Tsurumi}(2007)}]{wt2007}
\bibinfo{author}{\bibnamefont{Wakita}, \bibfnamefont{K.}}, and
  \bibinfo{author}{\bibfnamefont{T.}~\bibnamefont{Tsurumi}},
  \bibinfo{year}{2007}, in \emph{\bibinfo{booktitle}{WWW '07}}
  (\bibinfo{publisher}{ACM Press}), ISBN \bibinfo{isbn}{978-1-59593-654-7}.

\bibitem[{\citenamefont{Zachary}(1977)}]{z1977}
\bibinfo{author}{\bibnamefont{Zachary}, \bibfnamefont{W.~W.}},
  \bibinfo{year}{1977}, \bibinfo{journal}{J. Anthropological Res.}
  \textbf{\bibinfo{volume}{33}}, \bibinfo{pages}{452}.

\bibitem[{\citenamefont{Zakharov}(2007)}]{Zakharov07}
\bibinfo{author}{\bibnamefont{Zakharov}, \bibfnamefont{P.}},
  \bibinfo{year}{2007}, \bibinfo{journal}{Physica A}
  \textbf{\bibinfo{volume}{378}}(\bibinfo{number}{2}), \bibinfo{pages}{550}.

\end{thebibliography}
\bibliographystyle{apsrmp}

\end{document}